\documentclass[preprint,aps]{revtex4}
\usepackage{mathrsfs}
\usepackage{graphicx}

\begin{document}

\title{Electronic Evolution from the Parent Mott Insulator to a Superconductor in Lightly Hole-Doped Bi$_2$Sr$_2$CaCu$_2$O$_{8+\delta}$}


\author{Qiang Gao$^{1,2}$, Lin Zhao$^{1,**}$, Cheng Hu$^{1,2}$, Hongtao Yan$^{1,2}$, Hao Chen$^{1,2}$, Yongqing Cai$^{1,2}$, Cong Li$^{1,2}$, Ping Ai$^{1,2}$, Jing Liu$^{1,2,3}$, Jianwei Huang$^{1,2}$, Hongtao Rong$^{1,2}$, Chunyao Song$^{1,2}$, Chaohui Yin$^{1,2}$, Qingyan Wang$^{1}$, Yuan Huang$^{1}$,  Guodong Liu$^{1,2,4}$, Zuyan Xu$^{5}$ and X. J. Zhou$^{1,2,3,4,**}$}

\affiliation{
\\$^{1}$National Lab for Superconductivity, Beijing National laboratory for Condensed Matter Physics, Institute of Physics,
Chinese Academy of Sciences, Beijing 100190, China
\\$^{2}$School of Physical Sciences, University of Chinese Academy of Sciences, Beijing 100049, China
\\$^{3}$Beijing Academy of Quantum Information Sciences, Beijing 100193, China
\\$^{4}$Songshan Lake Materials Laboratory, Dongguan 523808, China
\\$^{5}$Technical Institute of Physics and Chemistry, Chinese Academy of Sciences, Beijing 100190, China
\\$^{**}$Corresponding author: LZhao@iphy.ac.cn, XJZhou@iphy.ac.cn
}



\maketitle

\newpage

{\bf High temperature superconductivity in cuprates is realized by doping the Mott insulator with charge carriers. A central issue is how such an insulating state can evolve into a conducting or superconducting state when charge carriers are introduced.  Here, by \textit{in situ} vacuum annealing and Rb deposition on the Bi$_2$Sr$_2$Ca$_{0.6}$Dy$_{0.4}$Cu$_2$O$_{8+\delta}$ (Bi2212) sample surface to push its doping level continuously from deeply underdoped ($T_c$=25 K, doping level \textit{p}$\sim$0.066) to the near zero doping  parent Mott insulator, angle-resolved photoemission spectroscopy measurements are carried out to observe the detailed electronic structure evolution in lightly hole-doped region for the first time. Our results indicate that the chemical potential lies at about l eV above the charge transfer band for the parent state at zero doping which is quite close to the upper Hubbard band. With increasing hole doping, the chemical potential moves continuously towards the charge transfer band and the band structure evolution exhibits a rigid band shift-like behavior.  When the chemical potential approaches the charge transfer band at a doping level of $\sim$0.05, the nodal spectral weight near the Fermi level increases, followed by the emergence of the coherent quasiparticle peak and the insulator-superconductor transition. Our observations provide key insights in understanding the insulator-superconductor transition in doping the parent cuprate compound and for establishing related theories.
}

The CuO$_{2}$ plane is a common and key structural element that dictates the unusual normal state and superconducting state properties in high temperature cuprate superconductors. Due to the strong electron-electron correlation, the undoped parent compound is a Mott insulator characterized by a charge transfer gap (CTG) between the charge transfer band (CTB) and the upper Hubbard band (UHB), derived from the hybridization of the lower Cu 3$d$ states and O 2$p$ states \cite{XGWen2006PALee,JWAllen1985JZaanen,LFMattheiss1987,TMRice1988FCZhang,FuggleJC1989FinkJ,YTokura2002NPArmitage}. How the insulating state of the parent compound evolves into a conductive and superconducting state upon electron or hole doping is a key issue in cuprate superconductors\cite{XGWen2006PALee,ZXShen2003ADamascelli,YTokura2002NPArmitage}. For the hole doped cuprates, two different doping evolution pictures have been proposed from angle-resolved photoemission spectroscopy (ARPES) measurements. In lightly doped La$ _{2-x} $Sr$ _{x}$CuO$ _{4} $, the chemical potential is found to be pinned and the in-gap states appear around the chemical potential in lightly hole-doped regime\cite{SUchida2000AIno,SUchida2003TYoshida,KKishio1997AIno,SUchida2006TYoshida}. As for Ca$_{2-x}$Na$_x$CuO$_2$Cl$_2$ (Na-CCOC)\cite{ITerasaki2003FRonning,ZXShen2004KMShen}, Bi$_2$Sr$_{2-x}$La$_x$CuO$_{6+\delta}$ (Bi2201) \cite{YAndo2008MHashimoto} and Bi$_2$Sr$_{2-x}$La$_x$(Ca,Y)Cu$_2$O$_{8+\delta}$ (Bi2212) \cite{ITerasaki2010KTanaka}, the chemical potential is shifted downward to the charge transfer band with increasing hole doping, and further doping continues to lower the chemical potential into the CTB with the emergence of coherent quasiparticle peaks. However, ARPES measurements on the detailed electronic evolution from the parent Mott insulator to  a superconductor, particularly in the lightly hole-doped regime ($p$=0$\sim$0.03), have been scarce so far.

Here we report, for the first time, ARPES measurements on the detailed electronic structure evolution from the parent Mott insulator to a superconductor  in Bi$_2$Sr$_2$CaCu$_2$O$_{8+\delta}$ (Bi2212) in the lightly hole-doped regime ($p$=0$\sim$0.07). Although Bi2212 has been studied extensively by ARPES in the last three decades, such measurements were not possible before because of the difficulty to get the undoped parent compound of Bi2212 and fine tuning of the hole concentration.  Instead of starting with the parent compound and then increasing the hole doping,  in this work,  we made it possible by starting with a heavily underdoped Dy-doped Bi$_2$Sr$_2$Ca$_{0.6}$Dy$_{0.4}$Cu$_2$O$_{8+\delta}$ (Dy-Bi2212) with a $T_c$ of 25 K and a hole doping level of 0.066, and then reducing the hole doping by \textit{in situ} vacuum annealing and Rb deposition on the sample surface during the ARPES measurements.  Compared with the usual way of chemical substitution in controlling the doping level,  the \textit{in situ} annealing and Rb deposition process have some obvious advantages . First, it can effectively avoid the effect of impurities which may induce different in-gap states at different doping levels  to affect the determination of chemical potential.  Second, it is easy to continuously control the hole concentration which is necessary to detect its detailed electronic structure evolution with doping. Third,  through the Rb deposition, it is possible to push the doping level to near zero, reaching the state of the parent Mott insulator. This is difficult  for the chemical substitution  in Bi$_2$Sr$_2$CuO$_6$ (Bi2201) and Bi2212 systems where 0.03 is the lowest doping level achieved so far\cite{YAndo2008MHashimoto,ITerasaki2010KTanaka}.

\begin{figure*}[tbp]
\begin{center}
\includegraphics[width=1.0\columnwidth,angle=0]{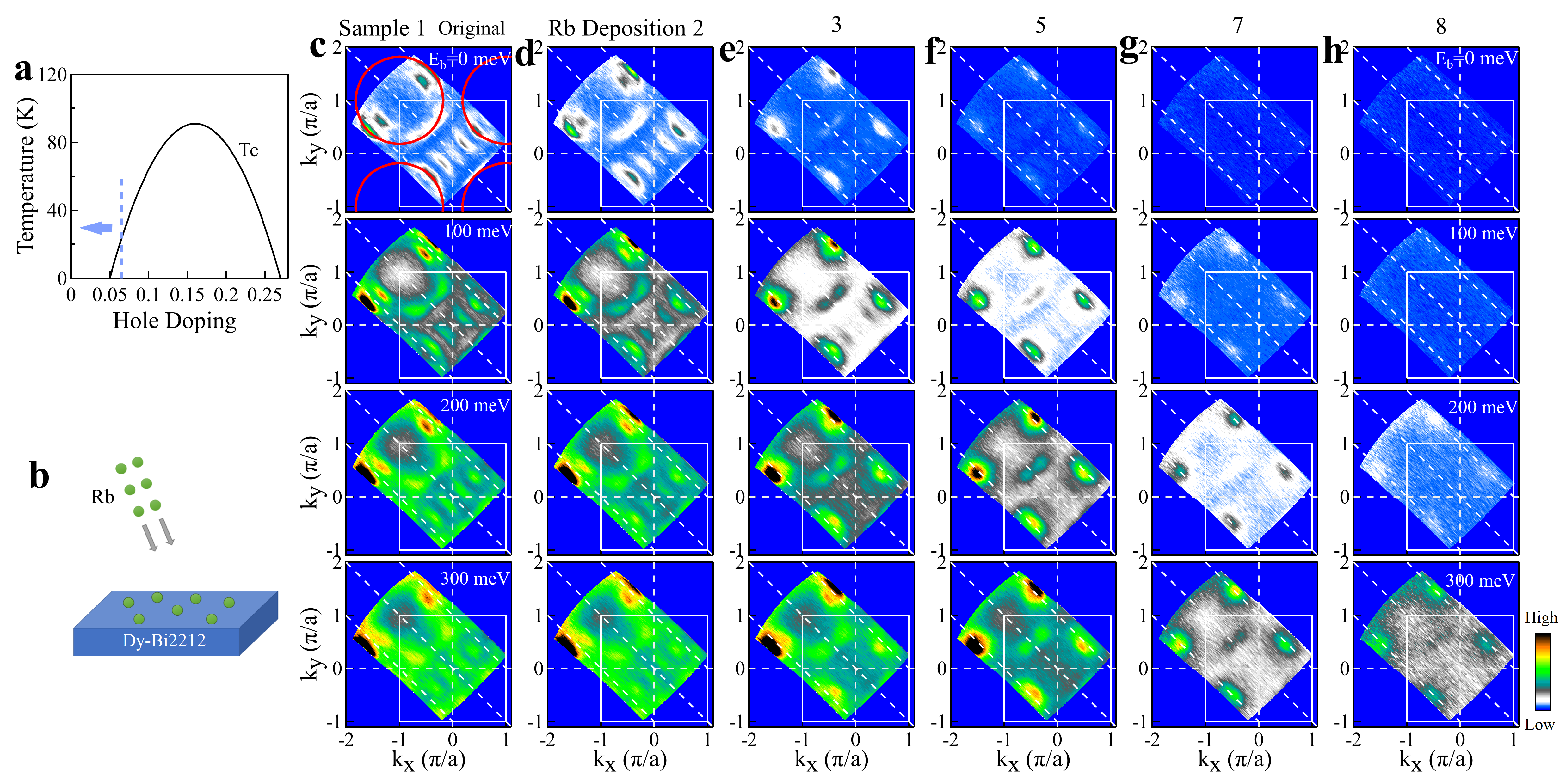}
\end{center}
\caption{{\bf  Doping evolution of Fermi surface and constant energy contours of Bi$_2$Sr$_2$Ca$_{0.6}$Dy$_{0.4}$Cu$_2$O$_{8+\delta}$ with Rb deposition.} (a) The variation of the superconducting transition temperature ($ T_c $) with the hole concentration (black line) in Bi2212.  The dashed blue line marks the initial hole concentration of the Dy-Bi2212 (\textit{$ T_c $}$\sim$25 K) sample before Rb deposition. Upon Rb deposition, the hole concentration of the sample surface decreases to approach zero, as marked by the blue arrow.  (b) Schematic illustration of \textit{in situ} Rb deposition onto the Dy-Bi2212 sample surface.  Electrons are transferred into the top layers due to the physical absorption of Rb on the sample surface. (c) The Fermi surface (top row) and constant energy contours at different binding energies of 100 meV (2nd row), 200 meV  (3rd row) and 300 meV (4th row) of the initial Dy-Bi2212 Sample 1 before Rb deposition. The images are obtained by integrating the spectral weight over an energy window of [-0.025, 0.025] eV with respect to different binding energies. The main Fermi surface is marked by the solid red lines. The white squares represent the first Brillouin zone. (d-h) Same as (c) but measured after different Rb deposition sequences.
}
\label{Fig1}
\end{figure*}

High quality Bi$_2$Sr$_2$Ca$_{0.6}$Dy$_{0.4}$Cu$_2$O$_{8+\delta}$ (Dy-Bi2212) single crystals were grown by the floating zone method and annealed in oxygen atmosphere.  The samples are heavily underdoped with a $T_c$ of 25\,K and corresponding hole concentration of $p$=0.066 as shown in Fig. \ref{Fig1}a. Angle-resolved photoemission measurements were performed on our lab-based system equipped with a Scienta DA30L analyzer and a Helium lamp with photon energies of 21.2 and 40.8 eV as the light source \cite{XJZhou2008GDLiu,XJZhou2018}. The overall energy resolution is set at 20 meV, and the angular resolution is $\sim$0.3$^\circ$. A SAES dispenser is used to evaporate rubidium (Rb) \textit{in situ} onto the Dy-Bi2212 surface step by step \cite{XJZhou2016YXZhangB} (Fig. \ref{Fig1}b). ARPES measurements were carried out after each Rb deposition and such a process is repeated many times. The samples can also be annealed in vacuum in the preparation chamber which is vacuum connected to the ARPES measurement chamber. All ARPES measurements were carried out at a temperature of 50\,K in the normal state, and in ultrahigh vacuum with a base pressure better than 5$\times$10$^{-11}$ Torr.

Figures \ref{Fig1}(c-h) show the evolution of the Fermi surface and constant energy contours at different binding energies of Bi$_2$Sr$_2$Ca$_{0.6}$Dy$_{0.4}$Cu$_2$O$_{8+\delta}$ with Rb deposition. For the initial Dy-Bi2212 sample before Rb deposition, the measured Fermi surface mapping (top panel in Fig. 1c) consists of the main Fermi surface (marked by red lines), weak superstructure Fermi surface and shadow Fermi surface \cite{KKadowaki2000HMFretwell}. It can also be seen that the Fermi surface spectral weight is mainly concentrated in the nodal region and heavily suppressed near the antinodal region, giving rise to a Fermi arc feature in the nodal region. Since the initial Dy-Bi2212 is heavily underdoped, the antinodal region is strongly suppressed due to the strong electron scattering and pseudogap formation. With Rb deposition on the Dy-Bi2212 sample surface, electrons are introduced that results in the decrease of the hole concentration. Figs. \ref{Fig1}(d-h) show the evolution of the Fermi surface and the constant energy contours for several typical Rb deposition sequences. With the increase of the Rb deposition, the Fermi arc length decreases gradually and  evolves into a small spot located at the ($\pi$/2,$\pi$/2) region for the deposition sequence 5 as shown in Fig. \ref{Fig1}f.  Further Rb deposition leads to the full suppression of the spectral weight at the Fermi level which is pushed to high binding energy (sequences 7 and 8 in Figs. \ref{Fig1}g and h). These results are consistent with previous observations in heavily underdoped cuprates\cite{ZXShen2006KTanaka,ITerasaki2010KTanaka,XJZhou2013YYPeng}.


 Figures 2a and 2b show corresponding band structures measured along two high symmetry momentum cuts crossing the nodal and antinodal points, respectively, on the Dy-Bi2212 sample measured in Fig. 1. For the initial sample before Rb deposition, the main band along the nodal direction (leftmost panel in Fig. 2a) crosses the Fermi level with a strong spectral weight near the Fermi level. On the other hand, the spectral weight near the antinodal region is strongly suppressed due to the formation of pseudogap in this heavily underdoped sample (leftmost panel in Fig. 2b). These results indicate that the physical properties in the heavily underdoped cuprates are mainly dictated by the nodal electron dynamics that will be the main focus of the work.  With the decrease of hole doping level upon Rb deposition, the spectral weight at the Fermi level gets suppressed along the nodal direction and disappears after the Rb sequence 3 (Fig. 2a),  indicating that the sample enters into an insulating state.  In the meantime,  the band structures exhibit an overall shift to higher binding energy both along the nodal direction (Fig. 2a) and near the antinodal region (Fig. 2b).  The band top moves $\sim$0.6 eV below the Fermi level for the Rb deposition sequence 8 (rightmost panel in Fig. 2a), which is similar to previous reports in Bi2201 and Bi2212 \cite{YAndo2008MHashimoto,ITerasaki2010KTanaka}.

\begin{figure*}[tbp]
\begin{center}
\includegraphics[width=1.0\columnwidth,angle=0]{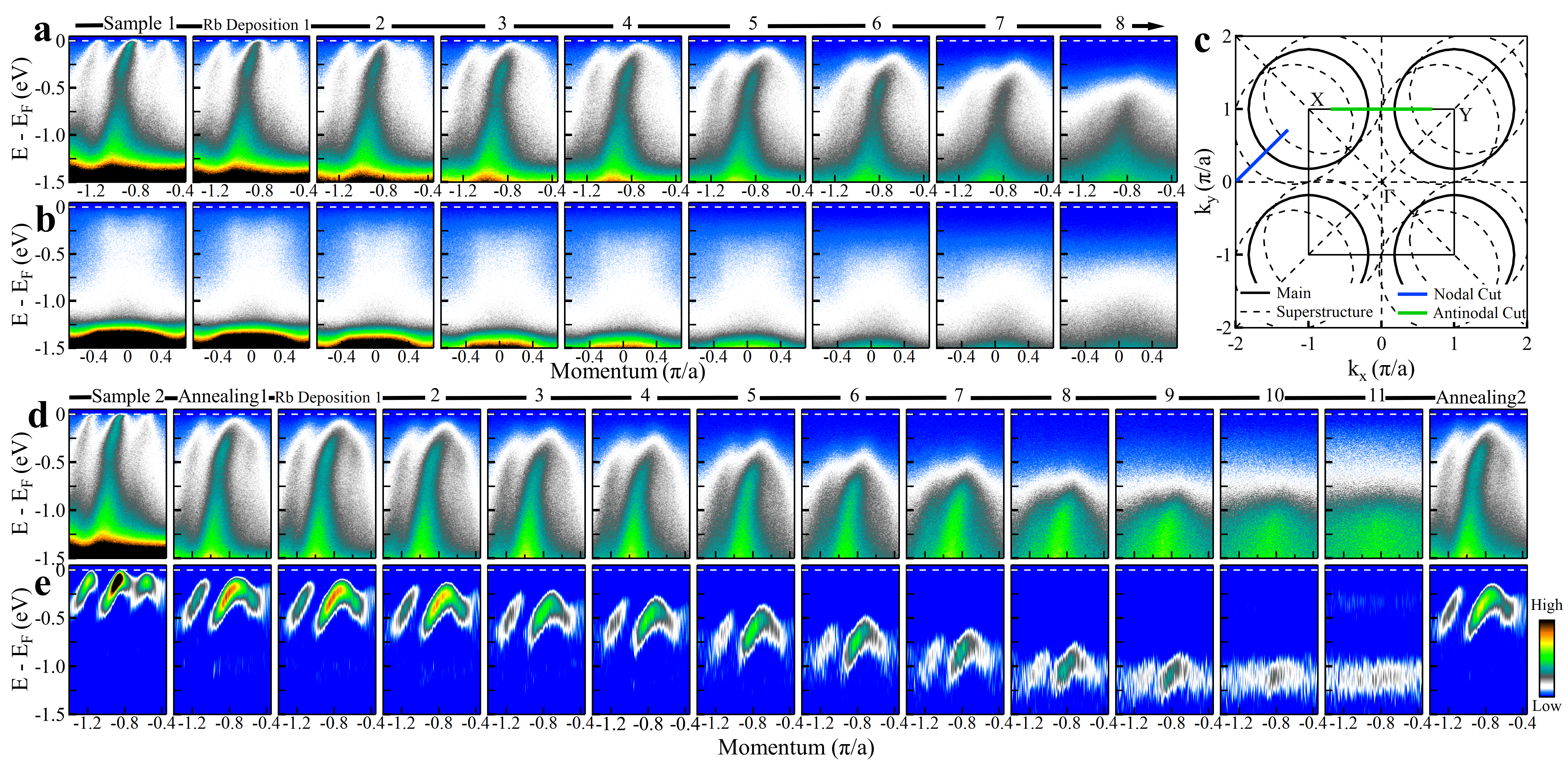}
\end{center}
\caption{{\bf Band structure evolution of Bi$_2$Sr$_2$Ca$_{0.6}$Dy$_{0.4}$Cu$_2$O$_{8+\delta}$ with Rb deposition.} (a) Doping evolution of band structure along the (-2$\pi$,0)-(-$\pi$,$\pi$) nodal momentum cut in the second Brillouin zone of the Sample 1.  (b) Doping evolution of band structure along the (-$\pi$,$\pi$)-($\pi$,$\pi$) antinodal momentum cut of the Sample 1.  (c) Schematic Fermi surface of Bi2212 consisting of the main Fermi surface (black solid lines) and first-order superstructure Fermi surface (black dashed lines). The locations of the nodal momentum cut for (a) and antinodal momentum cut for  (b) are marked by blue line and green line, respectively. (d) Doping evolution of band structure along the (-2$\pi$,0)-(-$\pi$,$\pi$) nodal momentum cut in the second Brillouin zone for the Sample 2.  For this measurement, the sample was first annealed in vacuum (column 2), and then followed by Rb deposition for 11 sequences (columns 3 to 13). After that, the sample was warmed up to remove Rb from the sample surface (right-most column)), and the band structure recovers to the initial state right after vacuum annealing (column 2).  (e) The second derivative images of (d) with respect to energy.
}
\label{Fig2}
\end{figure*}

In order to further enhance the reduction of  hole-doping level to approach zero, we first annealed another similar Dy-Bi2212 sample (Sample 2) in vacuum to remove extra oxygen, before we deposited Rb on the sample surface to further reduce the hole concentration. Fig. \ref{Fig2}d  shows the nodal band evolution with the vacuum annealing and Rb deposition. The corresponding second derivative images with respect to energy are shown in Fig. \ref{Fig2}e in order to determine the band top location. After vacuum annealing, the nodal main band already sinks below the Fermi level, indicating that it enters into an insulating state (the band for the annealing 1 in Fig. \ref{Fig2}d is similar to that for the Rb deposition sequence 3 in Fig. \ref{Fig2}a). With Rb deposition, the band keeps shifting to high binding energy, similar to the case in the Sample 1 (Fig. 2a).  But with the joint effect of vacuum annealing and Rb deposition, the nodal band shifts to nearly 1 eV below the Fermi level (Rb deposition sequence 9 in Figs. 2d and 2e) which is much deeper than $\sim$0.6 eV achieved in the sequence 8 of the Sample 1 (Fig. \ref{Fig2}a).  With further Rb deposition (sequences 10 and 11 in Figs. 2d and 2e), the band shift stops and appears to be pinned around 1 eV below the Fermi level. Considering that most cuprates (including Bi2212) can be doped by only one type of charge carrier, electrons or holes \cite{YAndo2006KSegawa,QKXue2019YZhong}, it is reasonable to think that the Sample 2  has reached a hole doping level close to zero. No further reduction of hole doping can occur even with more Rb deposition.  After the last Rb deposition (sequence 11), the sample was annealed again and the nodal band structure (rightmost panel annealing 2 in Figs. 2d and 2e) recovers to the case before Rb deposition (2nd leftmost panel annealing 1 in Figs. 2d and 2e). This indicates that Rb deposition is a physical adsorption process. It mainly introduces electrons into the sample and this deposition and  desorption process are reversible.

\begin{figure*}[tbp]
\begin{center}
\includegraphics[width=1.0\columnwidth,angle=0]{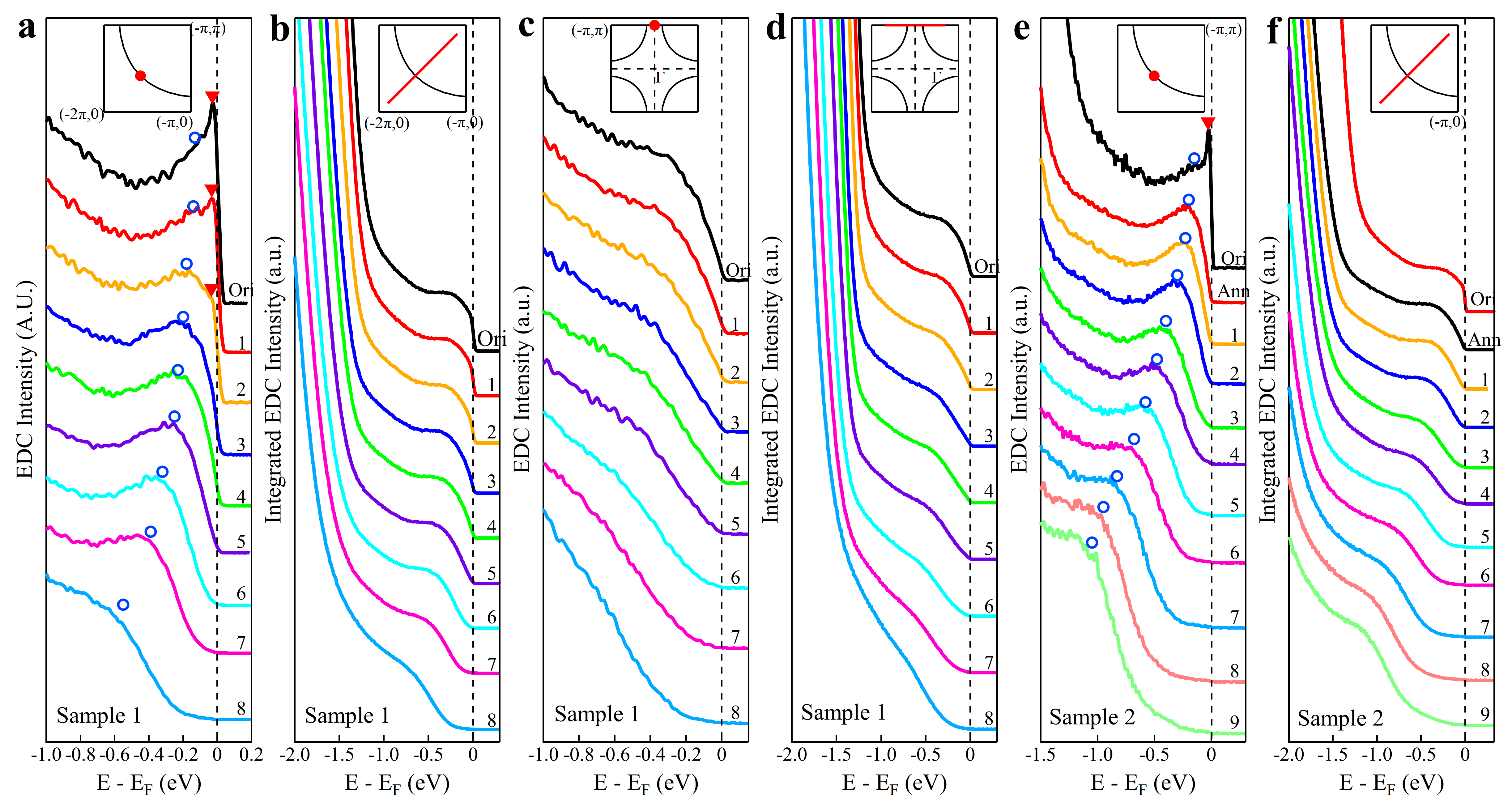}
\end{center}
\caption{{\bf Doping evolution of the photoemission spectra with the decreasing hole doping in Bi2212.}  (a) Doping evolution of energy distribution curves (EDCs) taken at the nodal point for the Sample 1.  A coherent peak can be observed in the initial sample and is marked by red triangle. The coherent peak gets weaker with Rb deposition, and the incoherent hump structure becomes prominent (marked by blue circles). With further Rb depositions, the incoherent peak shifts to high binding energy.  (b) Doping evolution of integrated EDCs taken along the nodal direction for the Sample 1.  The location of the nodal momentum cut is marked in the upper-right inset by the red line.  (c) Doping evolution of EDCs taken at (0,$\pi$) antinodal point for the Sample 1. (d) Doping evolution of integrated EDCs taken along the antinodal direction for the Sample 1. The location of the antinodal momentum cut is marked in the upper-right inset by the red line.  (e) Doping dependence of EDCs taken at the nodal point measured on the Sample 2. (f) Doping evolution of integrated EDCs taken along the nodal direction measured on the Sample 2. The location of the nodal momentum cut is marked in the upper-right inset by the red line.
}
\label{Fig3}
\end{figure*}

Figure \ref{Fig3} shows the  photoemission spectra (energy distribution curves, EDCs) in order to quantitatively track the electronic structure evolution with reducing hole doping in Dy-Bi2212 sample. Here we show both the angle-resolved photoemission spectra at the nodal point (Fig. 3a for the Sample 1 and 3e for the Sample 2) and antinodal point (Fig. 3c for the Sample 1), and the angle-integrated photoemission spectra along the nodal momentum cut (Fig. 3b for the Sample 1 and 3f for the Sample 2) and the antinodal cut (Fig. 3d for the Sample 1) which represent a local density of states.   Similar to the previous ARPES studies in hole-doped cuprates in the heavily underdoped region \cite{ZXShen2004KMShen,YAndo2008MHashimoto,ITerasaki2010KTanaka}, the nodal EDCs (Figs. 3a and 3e) mainly consist of two components for the initial sample with a doping level of \textit{p}$\sim$0.066, a sharp coherent peak near the Fermi level (marked by red triangles in Figs. 3a and 3e) and the broad incoherent hump at higher energy (marked by blue circles in Figs.  3a and 3e).  With the decrease of the hole doping, the coherent quasiparticle peak is suppressed rapidly, and the incoherent hump moves to high binding energy (Figs. 3a and 3e).  With the disappearance of the coherent quasiparticle peak, the leading edge of the EDCs starts to move away from the Fermi level,  indicating a transition into an insulating state.  The incoherent hump moves to $\sim$0.6 eV below the Fermi level in the Sample 1 (Fig. 3a) while it shifts to $\sim$1 eV with enhanced hole reduction in the Sample 2 (Fig. 3e).  We note that, while the incoherent hump structure lies well below the Fermi level, its spectral edge extends to a large range of energy, giving rise to some spectral weight even near the Fermi level. The integrated EDCs taken along the nodal direction in Fig. 3b and 3f also display a similar transition into an insulating state with the decrease of the hole doping level,  according to the position of the leading edge shift. On the other hand, near the antinodal region, no coherent peaks are observed in either angle-resolved spectra (Fig. 3c) or angle-integrated spectra (Fig. 3d). Only incoherent humps are observed that are all below the Fermi level, and shift to high binding energy with reducing hole doping.

 A key issue in doping a Mott insulator is how the chemical potential will shift which is associated with how the new states will emerge.  In hole-doped case, two possible routes are proposed \cite{XGWen2006PALee,RLGreene2010NPArmitage,MTachiki1989HMatsumoto, RLiu1990JWAllen,GASawatzky1993MBMeinders,WAGroen1994MAGroen, ZXShen2003ADamascelli}. One  is that the chemical potential is pinned in the middle of the gap, and the midgap states emerge at the chemical potential. The other is that the chemical potential will shift to the valence band upon hole doping.  To quantitatively determine the chemical potential shift, photoemission can measure the electronic states located far below the Fermi level so that their general spectral shapes remain qualitatively unchanged with doping, and their position change with doping can be used to extract the relative chemical potential shift \cite{ZXShen2004KMShen,YAndo2010MIkeda}.  In order to determine the chemical potential shift in Dy-Bi2212 sample with vacuum annealing and Rb deposition, we measured bismuth 5$d$ core level near the binding energy of $\sim$25 eV (Fig. 4a), and the high-binding energy O 2$p$ nonbonding states at $\Gamma$ (Fig. 4b) and at (0, $\pi$) point (Fig. 4c).  Keeping track on their relative peak position change,  a consistent result on the relative chemical potential shift with Rb deposition for the Sample 1 is obtained in Fig. 4d.   Fig.  4e shows the relative chemical potential shift obtained for the Sample 2 by similar procedure. In both of these two samples, the extracted chemical potential shift increases with vacuum annealing and Rb deposition, with an overall shift of $\sim$0.8 eV for the Sample 1 (Fig. 4d)  and $\sim$1.2 eV for the Sample 2 (Fig. 4e). These chemical potential shifts are consistent with the doping evolution of the band structure (Fig. 2) and EDCs (Fig. 3). Note that, for the deposition sequences 9-11 of the Sample 2, the chemical potential shift starts to become saturated, which is also consistent with the band structure evolution (Figs. 2d and 2e).
 
 In order to understand the evolution of electronic structure with doping, it is important to know the doping level corresponding to the vacuum annealing and each Rb deposition sequence. The systematic variation of the chemical potential with vacuum annealing and Rb deposition (Figs. 4d and 4e) provides us a good opportunity to estimate the doping level. First,  we notice that the chemical potential shift changes nearly linearly with the Rb deposition sequence number in both samples. Second, from the chemical potential shift, the vacuum annealing in the Sample 2 is equivalent to the Rb deposition sequence 3 in the Sample 1. Since each Rb deposition sequence is similar, for the Sample 2, it takes (3+9) times of Rb deposition to reduce the hole doping level from the initial \textit{p}$\sim$0.066 to nearly zero. This gives a hole-reducing efficiency for each Rb deposition, and thus makes it possible to estimate the doping level in Dy-Bi2212 after vacuum annealing and Rb depositions, as shown in Figs. 4f and 4g for the Sample 1 and Sample 2, respectively. In this way, all the previous electronic structure evolution (Figs. 1 and Fig. 2) with vacuum annealing and Rb deposition can be described on a quantitative doping level basis.

The doping induced electronic structure evolution (Figs. 2 and 3), combined with the determination of the chemical potential shift and the doping levels (Fig. 4), provides an electronic structure evolution picture in Bi2212 from the parent Mott insulator to a superconductor.
Fig. 5a shows the angle-resolved band structure measured along the nodal direction at several representative doping levels.  Different from Fig. 2 where the energy scale is referenced to the Fermi level, here the relative energy scale is referenced to the same high-binding energy position of the core level, thus highlighting the change of the chemical potential with doping.  In this case, it is straightforward to visualize the electronic structure variation with doping. The incoherent hump feature below the Fermi level observed in Figs. 2 and 3 along both the nodal and antinodal cuts represents the charge transfer band \cite{ITerasaki2003FRonning,XJZhou2018CHu,ZXShen2004KMShen,YAndo2008MHashimoto,ITerasaki2010KTanaka}. From our measurements, this charge transfer band is $\sim$1 eV below the chemical potential at near zero doping (bottom panel in Fig. 5a).  According to the tunneling measurements which can measure both the occupied state and unoccupied state, the charge transfer gap is about 1.2 eV in Bi2212 \cite{YYWang2016WRuan}.  This indicates, for undoped Bi2212,  the chemical potential locates close to the upper Hubbard band.  The proximity of the chemical potential to the upper Hubbard band in the parent compounds is also reported in other cuprates \cite{QKXue2019YZhong}.  This gives a schematic electronic structure of undoped Bi2212, as shown in the bottom panel of Fig. 5b. With increasing hole doping, the chemical potential gradually shift downwards to the charge transfer band.  With further hole doping to a level of $\sim$0.05, the chemical potential shifts into the charge transfer band, and the coherent quasiparticle peak emerges, and insulator-superconductor transition occurs. When the doping level further increases, the coherent peak gets sharper and the incoherent hump gets weaker.  We note that, since the upper Hubbard band always lies above the Fermi level in this hole-doping range, our ARPES measurements have no access to its evolution with the hole doping.


In summary, with \textit{in situ} tuning the hole concentration by vacuum annealing and Rb deposition to push the doping level continuously from deeply underdoped  to the near zero doping  parent Mott insulator, we have revealed the detailed electronic structure evolution of  Bi2212, for the first time,  from the parent Mott insulator to a superconductor.  Our results indicate that the chemical potential locates at about l eV above the charge transfer band within the charge transfer gap for the Bi2212 parent state. The chemical potential gradually shifts downwards to the charge transfer band in a  rigid band shift-like manner with the doping increase from 0 to $\sim$0.05.  When the chemical potential moves into the charge transfer band at \textit{p}$\sim$0.05, the nodal spectral weight near the Fermi level increases, followed by the emergence of the coherent quasiparticle peak and the insulator-superconductor transition. These observations provide key insights in understanding the insulator-superconductor transition in doping the parent cuprate compound and for establishing theories to understand the electronic evolution and emergence of superconductivity in high temperature cuprate superconductors.

\vspace{3mm}

\noindent {\bf Acknowledgement} We thank financial support from the National Natural Science Foundation of China (Grant Nos. 11888101, 11922414 and 11534007), the National Key Research and Development Program of China (Grant Nos. 2016YFA0300300 and 2017YFA0302900), the Strategic Priority Research Program (B) of the Chinese Academy of Sciences (Grant No. XDB25000000),  the Youth Innovation Promotion Association of CAS (Grant No.2017013), and the Research Program of Beijing Academy of Quantum Information Sciences (Grant No. Y18G06).

\vspace{3mm}

\noindent {\bf Author Contributions}\\
X.J.Z., L.Z. and Q.G. proposed and designed the research. Q.G. carried out the ARPES experiments with help from C.H., H.T.Y and H.C.. Q.G., H.C. and C.H.Y. prepared the Bi$_2$Sr$_2$Ca$_{0.6}$Dy$_{0.4}$Cu$_2$O$_{8+\delta}$ single crystals. Q.G., L.Z., C.H., H.T.Y., H.C., Y.Q.C., C.L., P.A., J.L., J.W.H., H.T.Y., C.Y.S., C.H.Y., Q.Y.W., Y.H., G.D.L, Z.Y.X. and X.J.Z. contributed to the development and maintenance of Laser-ARPES system. L.Z., Q.G. and X.J.Z. wrote the paper. All authors participated in discussions and comments on the paper.



\vspace{3mm}

\bibliographystyle{unsrt}

\newpage

\begin{figure*}[tbp]
\begin{center}
\includegraphics[width=1\columnwidth,angle=0]{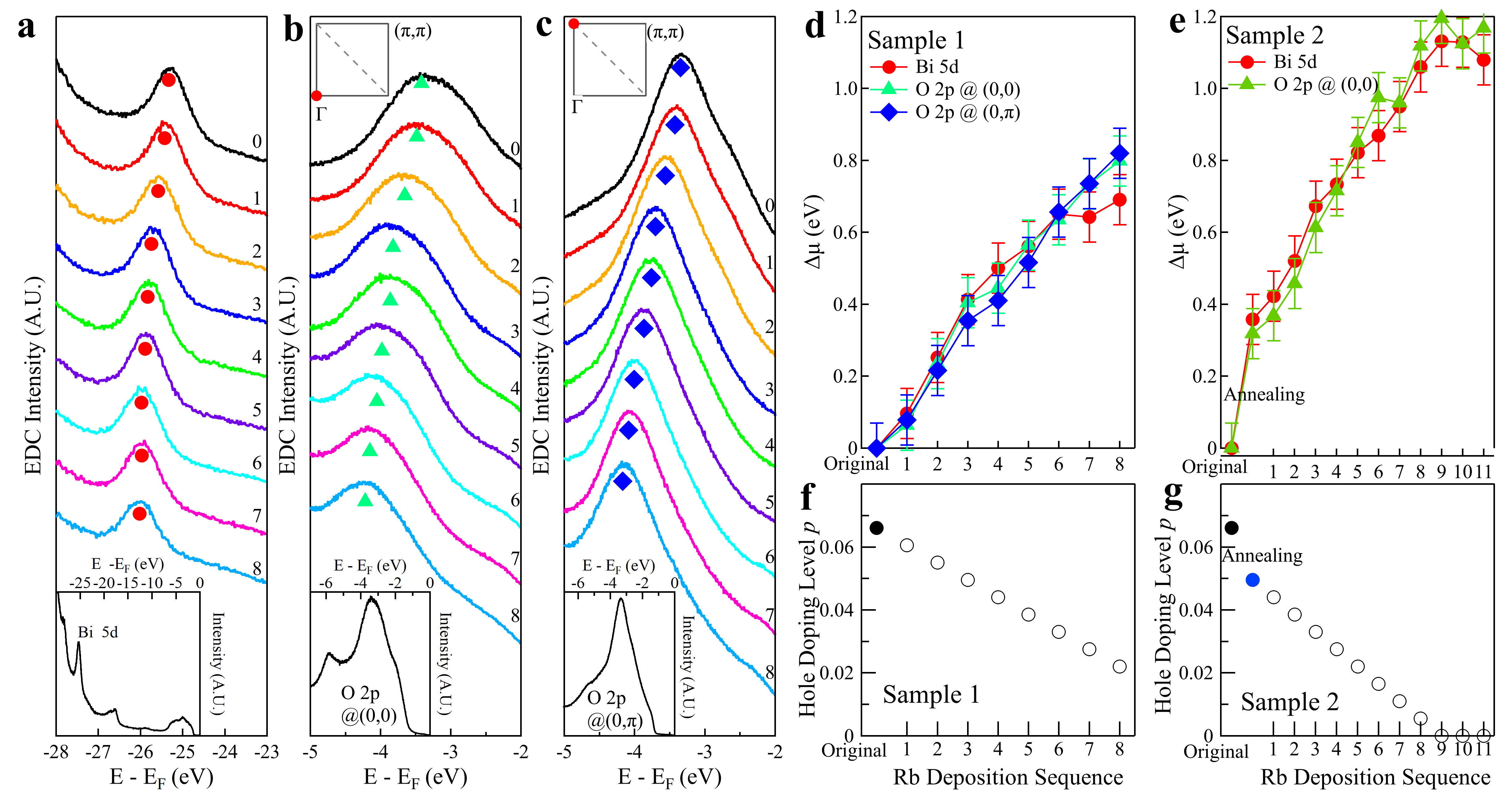}
\end{center}
\caption{{\bfseries Chemical potential shift with decreasing hole doping in Bi$_2$Sr$_2$Ca$_{0.6}$Dy$_{0.4}$Cu$_2$O$_{8+\delta}$.} (a-c) Photoemission spectra of the Bi 5$ d $ core level (a),  O 2$ p $ orbital at $\Gamma$ (b) and O 2$ p $ orbital at (0,$\pi$) (c) of the Sample 1 at various Rb deposition sequences. Symbols in (a-c) indicate the respective peak positions. (d) Chemical potential shift with Rb deposition for the Sample 1 determined from the three types of measurements in (a-c). The chemical potential shift for the original sample is set at zero as a reference. (e) Chemical potential shift with vacuum annealing and Rb depositions determined from the measurements of Bi 5$ d $ core level (red dots) and O 2$ p $ orbitals at $\Gamma$ (green triangles) for the Sample 2. (f) Estimated doping levels for various Rb depositions on the Sample 1.  (g) Estimated doping levels for vacuum annealing and various Rb depositions on the Sample 2.
}
\label{Fig4}
\end{figure*}

\begin{figure*}[tbp]
\begin{center}
\includegraphics[width=1\columnwidth,angle=0]{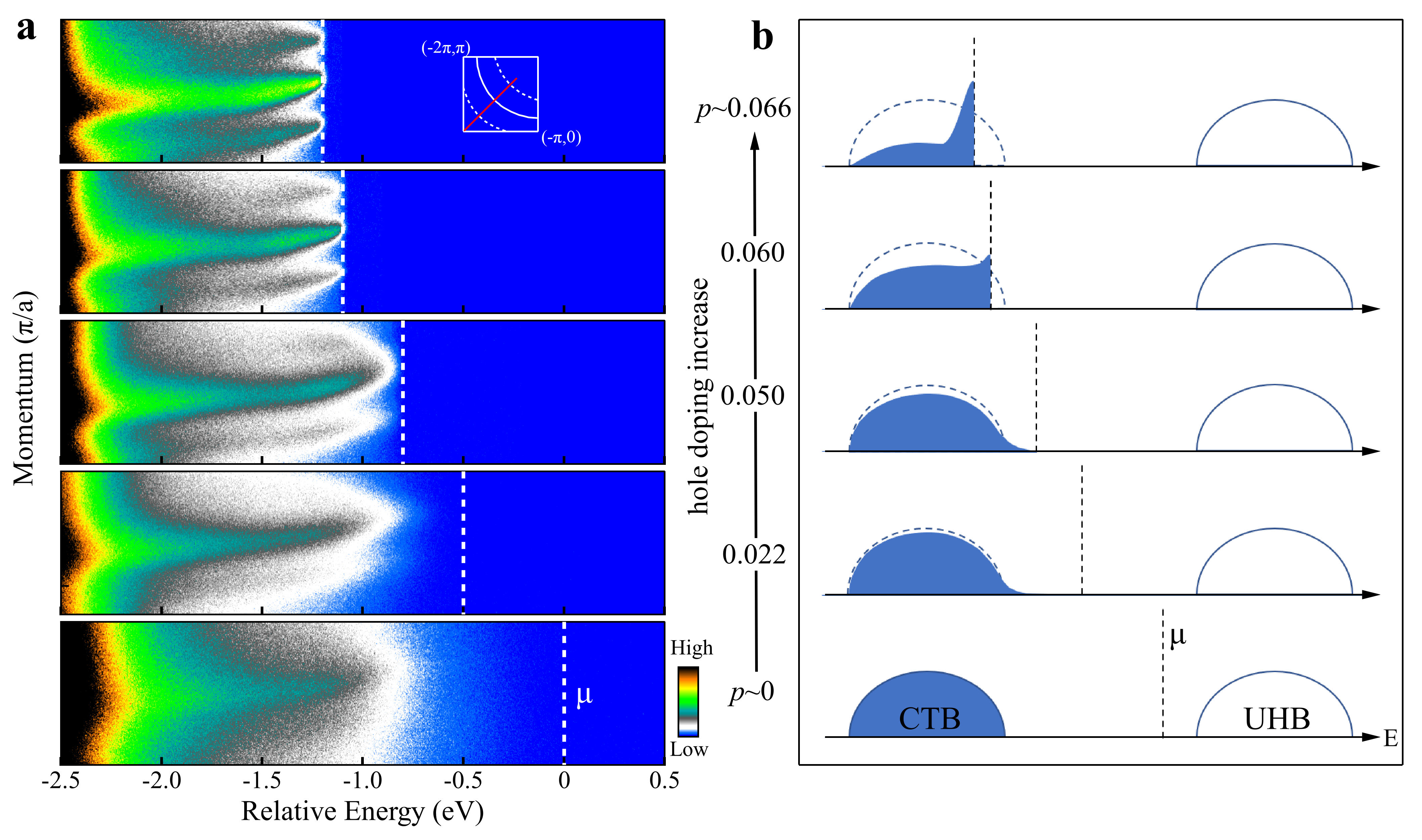}
\end{center}
\caption{{\bf Electronic structure evolution picture along the nodal direction of Bi$_2$Sr$_2$CaCu$_2$O$_{8+\delta}$ with hole doping from the parent Mott insulator to a superconductor.}  (a) Representative band structures at several representative hole doping levels, corresponding to the initial stage (\textit{p}$\sim$0) (bottom panel),   the Rb deposition sequence 1 of the Sample 1 (2nd panel from bottom),  the first annealing stage of the Sample 2  (3rd panel from bottom), and the Rb deposition stages of 5 (2nd panel from top)  and 9 (topmost panel) of the Sample 2, measured along the nodal direction in the second Brillouin zone. The horizontal axes are plotted on a relative energy scale where the chemical potential position of the zero doping is set at zero, and the energy position of others is referenced to the chemical potential shift $\mu$ shown in Fig. \ref{Fig4}.  The chemical potential positions for each doping are marked by white dashed lines. The corresponding momentum cut for the bands is shown in the upper-right inset. (b) Schematic electronic structure evolution with hole doping along the nodal direction in Bi2212 from a Mott insulator (bottom panel) to a superconductor (top panel, $p$$\sim$0.066).
}
\label{Fig5}
\end{figure*}

\end{document}